\documentclass[a4paper,fleqn]{cas-sc}
\usepackage[authoryear]{natbib}
\usepackage{graphicx,times}
\usepackage{subcaption}
\usepackage{amssymb,amsmath}
\usepackage{aas_macros}
\usepackage{hyperref}
\usepackage{comment}
\let\oldAA\AA
\renewcommand{\AA}{\text{\normalfont\oldAA}}
\usepackage{pdflscape}

\def\tsc#1{\csdef{#1}{\textsc{\lowercase{#1}}\xspace}}
\tsc{WGM}
\tsc{QE}

\begin{document}

\let\WriteBookmarks\relax
\def\floatpagepagefraction{1}
\def\textpagefraction{.001}

\shorttitle{Neutrino Association of PKS~0215+015}

\shortauthors{Bharathan et al.}

\title[mode = title]{Multi-wavelength Behaviour and Lepto-hadronic Modeling of PKS 0215+015 Around the Epoch of IceCube-220225A}

\author[1]{Athira M. Bharathan}
\cormark[1]
\ead{athirambharathan25@gmail.com}
\affiliation[1]{organization={Centre for Space Research, North-West University},
            city={Potchefstroom},
            country={South Africa}}

\author[1]{Markus B\"ottcher}

\cortext[1]{Corresponding author}

\begin{abstract}
The detection of high-energy astrophysical neutrinos has opened a new avenue for identifying cosmic particle accelerators. Among the candidate source populations, blazars, a class of active galactic nuclei (AGN) whose relativistic jets are oriented close to the observer's line of sight, have emerged as promising sites of hadronic interaction capable of producing PeV-scale neutrinos. In this work we investigate the flat-spectrum radio quasar (FSRQ) PKS~0215+015 ($z \simeq 1.715$), which was found to be positionally consistent with the \textit{IceCube-220225A} neutrino event of estimated energy $\sim$154~TeV. We present a multi-wavelength analysis combining \textit{Fermi}-LAT $\gamma$-ray data with \textit{Swift}-XRT and \textit{Swift}-UVOT observations to characterise the behaviour of the source around the neutrino detection epoch. PKS~0215+015 exhibited pronounced $\gamma$-ray, X-ray and optical activity during early 2022, consistent with the propagation of a disturbance along the jet. We construct broadband spectral energy distributions (SEDs) for the flaring interval coincident with the neutrino epoch and model these within leptonic and lepto-hadronic frameworks. The SED is well reproduced by a leptonic model (synchrotron, synchrotron self-Compton and external Compton emission), indicating that electron radiative processes dominate the observed emission. A co-accelerated proton population introduced to explore the maximum allowed hadronic contribution remains radiatively subdominant at all frequencies, but requires a proton kinetic luminosity roughly two orders of magnitude above equipartition with the magnetic field, illustrating the baryon-loading problem common to leptohadronic modeling of candidate neutrino blazars; the corresponding predicted neutrino flux falls well below the level implied by a single \textit{IceCube-220225A}-like detection. An independent analysis of the short-term $\gamma$-ray variability yields an emission-region size consistent with that obtained from the SED fit, supporting the physical self-consistency of the adopted model. We discuss the temporal and energetic plausibility of a physical association between PKS~0215+015 and \textit{IceCube-220225A} and consider the implications for future multimessenger studies of high-redshift FSRQs.
\end{abstract}


\maketitle

\section{Introduction}
\label{sec:intro}

The IceCube Observatory's detection of high-energy astrophysical neutrinos has opened a new window onto the most energetic processes in the Universe \citep{2018Sci...361..147I}. The clearest evidence that blazar jets can produce these neutrinos came in 2017, when a 290~TeV neutrino was found to coincide with a flare of the blazar TXS~0506+056, showing that relativistic AGN jets can accelerate hadrons to energies high enough for photohadronic interactions \citep{2018Sci...361.1378I}. This result established blazars as leading candidates for high-energy neutrino production and triggered a systematic search for further associations.

Since then, several IceCube alerts have been found to coincide, in both position and time, with bright blazars \citep{2018Sci...361.1378I, 2014PhRvL.113j1101A, 2019ATel12967....1T, 2024GCN.35485....1I}. One such case is \textit{IceCube-220225A}, a neutrino event detected on 25~February~2022 with a reconstructed energy of $\sim$150~TeV, which was found to be spatially consistent with the FSRQ PKS~0215+015 at $z = 1.715$ \citep{1987AJ.....93..529F, 2022GCN.31650....1I}.

PKS~0215+015 has an interesting classification history that bears directly on its plausibility as a neutrino source. It was originally classified as a BL~Lacertae (BL~Lac) object based on early spectrophotometric and polarimetric observations that showed weak or absent emission lines \citep{1982ApJ...252..447G}, but later observations revealed prominent broad emission lines, leading to its reclassification as an FSRQ \citep{1982MNRAS.200.1091B, 1987AJ.....93..529F}. This kind of transition, where a featureless BL~Lac-like continuum during low states gives way to an FSRQ-like broad-line spectrum during brighter, high-accretion states, is a recognised pattern among changing-look blazars \citep{2023NatAs...7.1282R}. It matters here because FSRQs are considered efficient neutrino producers precisely because of this broad-line region (BLR): the dense BLR and accretion-disc photon fields provide abundant target photons for the photohadronic ($p\gamma$) interactions needed to make PeV-scale neutrinos \citep{2001PhRvL..87v1102A, 2014PhRvD..90b3007M, 2016NatPh..12..807K}. If PKS~0215+015 was in a more FSRQ-like, BLR-dominated state around the time of the neutrino detection, that would strengthen the case for a physical association.

Multi-wavelength monitoring around the neutrino epoch shows the source in an active state across several wavebands. \textit{Fermi}-LAT recorded several pronounced $\gamma$-ray flares between 2021 and 2023, radio monitoring from OVRO and VLBI observations showed enhanced activity and changing polarization consistent with a disturbance moving along the jet, and a VLBI campaign reported the emergence of a new superluminal jet component close to the neutrino detection epoch \citep{2026A&A...708A.326E}. An independent radio campaign also reported a major outburst beginning in mid-2021, with flux densities exceeding 3~Jy, the highest radio state recorded for this source in more than two decades \citep{2022ATel15247....1P}. Taken together, these observations point to a sustained period of enhanced jet activity bracketing the neutrino detection.

This kind of coincidence is worth taking seriously because of what it could reveal about the underlying emission mechanism. The broadband SED of a blazar is made up of two components: a low-energy component from radio to optical/X-ray frequencies, attributed to synchrotron radiation from relativistic electrons, and a high-energy component whose origin is still debated. In leptonic models, this high-energy emission comes from inverse-Compton scattering of synchrotron or external photons by the same electrons; in hadronic models, it comes from proton synchrotron radiation and electromagnetic cascades triggered by photohadronic interactions. If hadrons are accelerated alongside electrons in the jet, $p\gamma$ interactions with either the jet's own synchrotron photons or an external radiation field such as the BLR can generate PeV-scale neutrinos \citep{1993A&A...269...67M, 2014PhRvD..90b3007M}. Because neutrinos arise naturally from hadronic interactions but not from purely leptonic ones, candidate neutrino blazars like PKS~0215+015 offer a direct test of which mechanism dominates.

Confirming a physical connection between a single neutrino event and a blazar flare is not straightforward, however, and requires a multi-pronged approach: the positional and temporal coincidence must be weighed against the IceCube localisation uncertainty and the source's typical variability; the multi-wavelength activity must be checked for conditions favourable to neutrino production; and broadband SED modelling is needed to constrain the jet parameters, particle distributions and baryon loading required to reproduce both the photon and neutrino fluxes together.

In this work we carry out exactly this kind of analysis for PKS~0215+015 in the context of the \textit{IceCube-220225A} alert. Using \textit{Fermi}-LAT $\gamma$-ray data together with \textit{Swift}-XRT and \textit{Swift}-UVOT observations, we characterise the source's variability across timescales and wavebands, construct time-resolved SEDs for the flaring state coincident with the neutrino epoch, and apply both leptonic and lepto-hadronic models to explore what emission scenarios are consistent with the data. A short-term variability analysis is used to independently constrain the size of the emission region. We then discuss the physical and statistical plausibility of an association with \textit{IceCube-220225A}, with a view to identifying observational signatures that can guide future multimessenger searches.

\section{Observations and Data Analysis}
\label{sec:data}

\subsection{\textit{Fermi}-LAT}

To reduce the data from the Large Area Telescope (LAT; \citealt{2009ApJ...697.1071A}) on board \textit{Fermi}, we used the \textsc{fermipy} package \citep{2017ICRC...35..824W}. We analysed data in the energy range 100~MeV to 300~GeV, considering all SOURCE-class events (\texttt{evclass}=128, \texttt{evtype}=3) within a $15^{\circ}$ region of interest centred on the source. To ensure data quality we applied the standard filter \texttt{DATA\_QUAL>0 \&\& LAT\_CONFIG==1}. For the background models we used the Galactic diffuse emission model (\texttt{gll\_iem\_v06}) and the isotropic diffuse emission model (\texttt{iso\_P8R2\_SOURCE\_V6\_v06}). A binned likelihood analysis was performed to generate the $\gamma$-ray spectra and light curves. For the SED analysis of PKS~0215+015 we used data covering the interval 15~February~2022 (MJD~59625) to  07~March~2022 (MJD~59645), corresponding to ten days before and after the neutrino detection (MJD~59635).

\subsection{IceCube Event IC220225A}
\label{sec:icecube}

The \textit{IceCube-220225A} event was detected on 25~February~2022 at 14:12:00.7~UT and reported by the IceCube Collaboration via the Gamma-ray Coordinates Network (GCN 31650). The event was selected by the \texttt{ICECUBE\_Astrotrack\_Bronze} alert stream; Bronze-tier alerts have an average astrophysical neutrino purity of 30\%, and this specific alert had an estimated atmospheric false-alarm rate of 2.329~yr$^{-1}$. Following offline reconstruction, the best-fit equatorial coordinates were refined to $\mathrm{RA} = 34.7^{+3.1}_{-2.6}$~deg, $\mathrm{Dec} = 0.0^{+1.8}_{-1.5}$~deg (J2000, 90\% PSF containment.

The same circular identified the bright FSRQ PKS~0215+015 (4FGL~J0217.8+0144, RA~=~34.46~deg, Dec~=~1.73~deg J2000) as lying $\sim$1.75~deg from the neutrino best-fit position, the only source in both the 4FGL-DR2 and 3FHL catalogues located within the 90\% containment region. A dedicated \textit{Fermi}-LAT analysis confirmed this association and reported that PKS~0215+015 had been in an enhanced $\gamma$-ray state since mid-2021, with a flux (E$>$100~MeV) in the day preceding the neutrino detection more than seven times the 4FGL-DR3 average, and more than three times the average when integrated over the preceding month \citep{2022ATel15243....1G}.
No other prominent $\gamma$-ray emitter within the 90\% error circle is listed in the \textit{Fermi}-LAT 4FGL catalogue \citep{2020ApJS..247...33A}, further strengthening the case for a potential association.

No additional neutrino events spatially coincident with PKS~0215+015 have been reported in public IceCube analyses or time-dependent searches to date. Nevertheless, the positional consistency and contemporaneous multi-wavelength activity make \textit{IceCube-220225A} a valuable case for examining the physical conditions under which FSRQs may produce high-energy neutrinos.

\subsection{\textit{Swift}-XRT}

We used data in the 0.3--10~keV band from the \textit{Swift} X-ray Telescope (XRT; \citealt{2005SSRv..120..165B}), obtained from the HEASARC archive.\footnote{\url{https://heasarc.gsfc.nasa.gov/docs/archive.html}} Data acquired in photon-counting mode were used for both timing and spectral analysis, and were reduced using \texttt{xrtpipeline} with the calibration files from HEASOFT~v6.30 and default parameters recommended by the instrument team. Source spectra were extracted from a circular region of 60" radius, and background spectra from an adjacent source-free circular region of 120" radius. Exposure maps were generated with \textsc{ximage}, and ancillary response files with \texttt{xrtmkarf}. Spectral fitting in \textsc{xspec} employed an absorbed power-law model, with the Galactic neutral hydrogen column density fixed at $N_{\rm H} = 3.35\times10^{20}$~cm$^{-2}$ towards PKS~0215+015 \citep{2016A&A...594A.116H}.

\subsection{\textit{Swift}-UVOT}

Ultraviolet and optical photometry was obtained with the \textit{Swift} Ultraviolet/Optical Telescope (UVOT; \citealt{2005SSRv..120...95R}) on board the \textit{Swift} spacecraft \citep{2004ApJ...611.1005G}. Observations in the $V$, $U$, $UVW1$ and $UVW2$ filters were used, with central wavelengths (full width at half maximum) of 5468~\AA\ (796~\AA), 3465~\AA\ (785~\AA), 2600~\AA\ (693~\AA) and 1928~\AA\ (657~\AA), respectively \citep{2008MNRAS.383..627P}. Data were processed using the standard online tools of the \textit{Swift} data archive, which perform calibration and photometric extraction automatically. Observed fluxes were corrected for Galactic extinction using the appropriate values for the line of sight to the source \citep{2011ApJ...737..103S}.

\section{Temporal and Spectral Analysis}
\label{sec:results}

\subsection{multi-wavelength Light Curves}

\begin{figure*}
\centering
\includegraphics[width=0.95\textwidth]{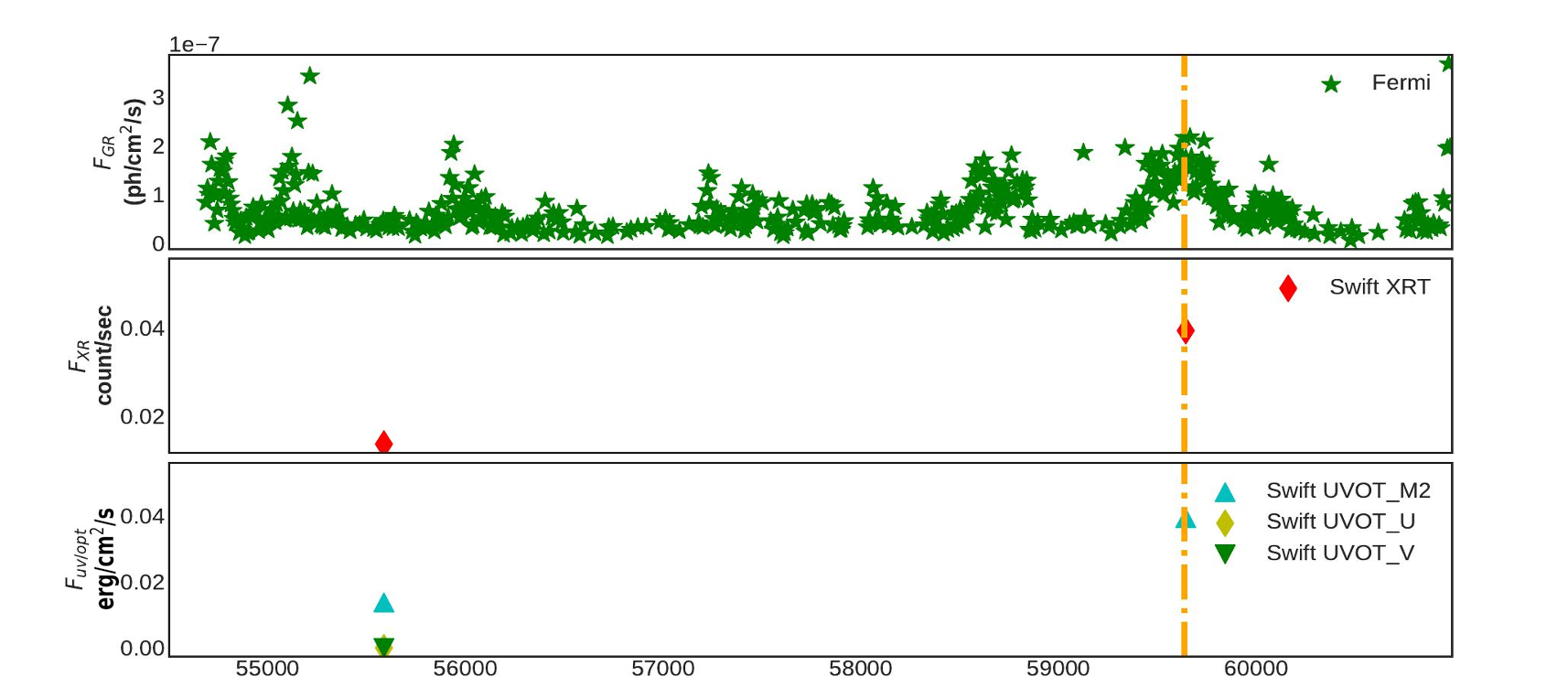}
\caption{Multi-wavelength light curve of PKS~0215+015 spanning 18 years (MJD~54749-61170). From top to bottom: the weekly binned \textit{Fermi}-LAT $\gamma$-ray light curve (100~MeV--300~GeV); the \textit{Swift}-XRT count rate obtained in photon-counting mode; and the \textit{Swift}-UVOT light curves in the $UVM2$, $U$ and $V$ bands. The vertical dashed line marks the epoch of the \textit{IceCube-220225A} neutrino detection (MJD~59635).}
\label{figure-1}
\end{figure*}

The long-term light curve of a blazar traces its flux evolution across the electromagnetic spectrum and is essential for identifying quiescent and active states, locating the epoch of interest within the source's broader variability history, and establishing the multi-wavelength context needed to interpret any single flaring episode, such as the one coincident with a neutrino detection. To this end, we constructed the multi-wavelength light curve of PKS~0215+015 spanning approximately 18 years, from 2008~August to May 2026 (MJD~54749 to MJD~61170), comprising the weekly binned \textit{Fermi}-LAT $\gamma$-ray flux, the \textit{Swift}-XRT count rate, and the \textit{Swift}-UVOT photometry in the $UVM2$, $U$ and $V$ bands, shown in Figure~\ref{figure-1}. The vertical dashed line marks the epoch of the \textit{IceCube-220225A} detection (MJD~59635). Over this baseline, PKS~0215+015 alternates between extended quiescent intervals and episodes of pronounced multi-wavelength activity, with a clear enhancement in the $\gamma$-ray and optical/UV bands coincident with the neutrino detection epoch.

\subsection{Short-Term Variability and Emission Region Size}
\label{sec:variability}

Blazar flux variability provides an independent, model-free constraint on the physical size of the emitting region through simple light-crossing arguments. Blazars are known to show flux variations on timescales shorter than an hour \citep{2011A&A...530A..77F, 2013ApJ...766L..11S, 2022A&A...668A.152P}. 
We quantify the shortest significant variations in the \textit{Fermi}-LAT light curve around the neutrino epoch using the flux-doubling (or halving) timescale, defined for two flux measurements $F(t_{0})$ and $F(t)$ at times $t_0$ and $t$ ($t>t_0$) as
\begin{equation}
\label{eq:tdouble}
F(t) = F(t_{0})\times2^{-(t-t_{0})/\tau},
\end{equation}
requiring the flux difference between the two epochs to exceed $3\sigma$ \citep{2011A&A...530A..77F}. The shortest doubling/halving timescale identified in the light curve during the flaring interval bracketing MJD~59635 is $\tau \simeq 6.61$~hrs.

The limit to the size of the emission region follows from the requirement that the region be able to communicate a change in flux across its own light-crossing time, corrected for relativistic beaming and cosmological redshift:
\begin{equation}
\label{eq:Rsize}
R \lesssim \frac{c\,\delta\,\tau}{1+z},
\end{equation}
where $\delta$ is the Doppler factor of the emitting region and $z = 1.715$ is the redshift of PKS~0215+015. Adopting the Doppler factor obtained from the broadband SED fit (Section~\ref{sec:sedmodel}), $\delta = 20$, Equation~\ref{eq:Rsize} yields an emission region size of $R \lesssim$ 5.17$\times$ 10$^{15}$~cm.

This derived size is compared with the value obtained independently from the SED modeling in Section~\ref{sec:sedmodel}, which is found to be 5.0 $\times$ 10$^{15}$ cm. The consistency between the two provides a check on the physical self-consistency of the adopted model parameters.

\section{Broadband SED modeling}
\label{sec:sedmodel}

We model the broadband SED constructed for the flaring epoch coincident with \textit{IceCube-220225A} using the time-independent, one-zone leptonic and lepto-hadronic framework of \citet{2013ApJ...768...54B}. In both cases, the emission region is taken to be a spherical blob of radius $R$, filled with a tangled magnetic field $B$, moving with bulk Lorentz factor $\Gamma$ and Doppler factor $\delta$ along the jet.

\subsection{Leptonic Model}
\label{sec:leptonic}

Relativistic electrons are injected into the region with an energy distribution given by a power law with an exponential cutoff.
The steady-state electron distribution is obtained by solving the kinetic equation, including radiative cooling (synchrotron and inverse-Compton) and particle escape on a timescale $t_{\rm esc} = \eta_{\rm esc}R/c$. The low-energy SED component (radio to optical/X-ray) is produced by synchrotron radiation from this electron population, while the high-energy component (X-ray to $\gamma$-ray) is produced by inverse-Compton scattering. We include synchrotron self-Compton (SSC), in which the electrons up-scatter their own locally produced synchrotron photons, together with two distinct external-Compton (EC) channels. The first is EC scattering of the direct, anisotropic radiation field of the accretion disc, computed self-consistently from the central black hole mass and the accretion-disc luminosity $L_{\rm disc}$. The second is EC scattering of an isotropic external radiation field, representative of reprocessed emission from the broad-line region (BLR) or the dusty torus, approximated in the co-moving frame as a blackbody of temperature $T_{\rm ext}$ and energy density $u_{\rm ext}$; this isotropic field is parameterised independently of the disc-EC component. The model provides a self-consistency check through the minimum variability timescale, $t_{\rm var} = (1+z)R/(\delta c)$, which can be compared directly with the observed short-term variability.

Applying this framework to the flaring epoch coincident with \textit{IceCube-220225A}, the SED is reproduced with an emission region of radius $R = 5\times10^{15}$~cm and magnetic field $B = 1$~G, moving with bulk Lorentz factor $\Gamma = 20$ at a viewing angle $\theta_{\rm obs} = 2.866^{\circ} \simeq 1/\Gamma$, giving a Doppler factor $\delta \simeq \Gamma = 20$. The electron population is injected with a minimum Lorentz factor $\gamma_{\min} = 1.2\times10^{5}$, a cutoff at $\gamma_{c} = 1\times10^{6}$, and an injection spectral index $q = 5.5$, with $\eta_{\rm esc} = 1$. Both of these values are unusual: $\gamma_{\min}$ is very high compared to the values typically invoked in one-zone leptonic blazar models, and is difficult to justify without invoking a selective, high-threshold acceleration mechanism rather than standard shock or turbulent acceleration; similarly, $q = 5.5$ is considerably steeper than the $q \sim 2$--$3$ expected from standard first-order Fermi acceleration. We discuss the physical implications of this narrow, high-energy electron distribution further in Section~\ref{sec:discussion}.

The disc-EC component is computed from a central black hole mass of $5\times10^{8}\,M_{\odot}$ and $L_{\rm disc} = 1\times10^{46}$~erg~s$^{-1}$; the isotropic (BLR) external radiation field is set independently, characterised by $T_{\rm ext} = 1.0\times10^{4}$~K and $u_{\rm ext} = 8.0\times10^{-2}$~erg~cm$^{-3}$.
This configuration implies a minimum variability timescale of $t_{\rm var} = (1+z)R/(\delta c) \simeq 6.3$~hr, broadly consistent with the short-term $\gamma$-ray variability measured directly from the light curve.

The resulting electron kinetic luminosity is $L_{e} = 5.8\times10^{43}$~erg~s$^{-1}$ and the Poynting-flux (magnetic) luminosity is $L_{B} = 3.75\times10^{43}$~erg~s$^{-1}$, giving $L_{B}/L_{e} \simeq 0.65$: close to equipartition, and indicative of a physically self-consistent jet configuration. We find that this leptonic (synchrotron~+~SSC~+~EC) model reproduces the observed broadband SED, from optical through $\gamma$-ray energies; the corresponding parameters are listed in Table~\ref{tab:model_parameters}.

\subsection{Hadronic Contribution}
\label{sec:hadronic}

Although the broadband SED is satisfactorily reproduced by the purely leptonic model of Section~\ref{sec:leptonic}, the spatial and temporal coincidence of the flare with \textit{IceCube-220225A} motivates a check for a possible hadronic contribution, since leptonic processes alone cannot account for the production of high-energy neutrinos. To evaluate the maximum hadronic contribution and associated neutrino flux consistent with the data, we extended the leptonic model of Section~\ref{sec:leptonic} to include a co-accelerated relativistic proton population, injected with the same functional form adopted for the electrons, a power law with an exponential cutoff. The protons radiate via proton synchrotron emission and via photohadronic ($p\gamma$) and the isotropic external radiation field. The $p\gamma$ interactions are treated with the semi-analytical formalism of \citet{2013ApJ...768...54B}, based on the interaction templates of \citet{2008PhRvD..78c4013K}, which provide the resulting spectra of secondary photons, neutrinos and electron--positron pairs. The model self-consistently accounts for secondary emission, including photons from $\pi^{0}$ decay, synchrotron radiation from secondary $e^{\pm}$ produced in charged pion and muon decay, and synchrotron-supported pair cascades initiated by internal $\gamma\gamma$ absorption.

\begin{figure*}
\centering
\includegraphics[width=0.95\textwidth]{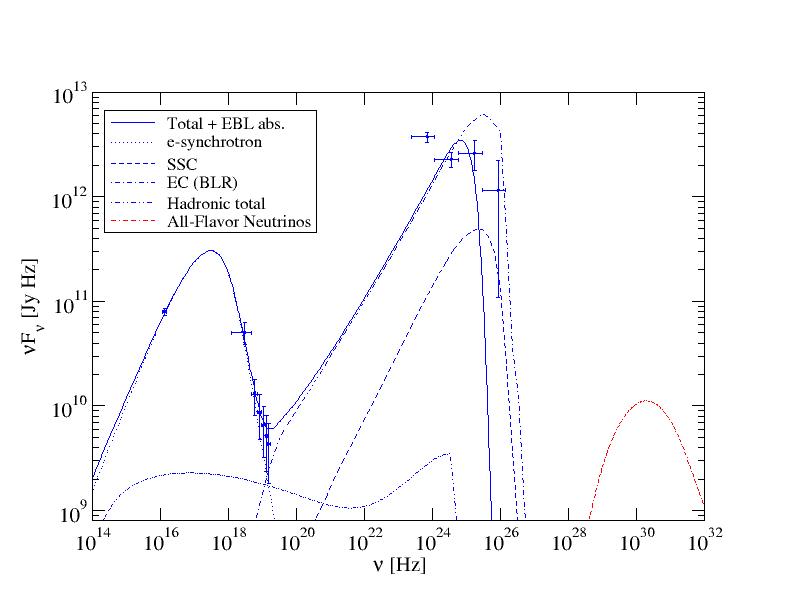}
\caption{Broadband spectral energy distribution of PKS~0215+015 during the flaring epoch coincident with \textit{IceCube-220225A}. Multiwavelength data (optical/UV, X-ray and $\gamma$-ray) are shown together with the best-fit leptonic (synchrotron and SSC) and hadronic (proton synchrotron and photo-hadronic cascade) model components, computed using the framework of \citet{2013ApJ...768...54B}. The predicted all-flavour photohadronic neutrino spectral energy distribution corresponding to the best-fit hadronic model is also shown with pink dotted line.}
\label{figure-3}
\end{figure*}

Applying this framework to the flaring epoch of PKS~0215+015 coincident with \textit{IceCube-220225A}, and adopting the same emission-region geometry as the leptonic fit ($R$, $B$, $\Gamma$, $\delta$; Section~\ref{sec:leptonic}), we inject protons with a spectral index $\alpha_{p} = 2.0$ between $E_{p,\min} = 1$~GeV and $E_{p,\max} = 10^{10}$~GeV, corresponding to a maximum allowed proton kinetic luminosity of $L_{p} = 4\times10^{45}$~erg~s$^{-1}$ to remain consistent with the observed SED. This corresponds to $L_{B}/L_{p} \simeq 9.4\times10^{-3}$ and $L_{e}/L_{p} \simeq 1.5\times10^{-2}$, i.e.\ the proton population could carry roughly 70 times more kinetic power than the radiating electrons, yet, owing to the inherent inefficiency of proton synchrotron and $p\gamma$ radiative channels, this large energy reservoir still produces a hadronic (proton synchrotron and $p\gamma$-cascade) contribution that remains subdominant to the leptonic (synchrotron/SSC/EC) emission at all frequencies (Fig.~\ref{figure-3}). The corresponding predicted all-flavour neutrino flux (Fig.~\ref{figure-3}) is correspondingly low, and, given the effective area of IceCube at the relevant energy, $E_{\nu} \simeq 154$~TeV, implies an expected neutrino event rate well below one detection over the lifetime of IceCube. This large disparity between the proton and electron kinetic luminosities represents a significant super-equipartition baryon loading. The best-fit hadronic model parameters are listed alongside the leptonic parameters in Table~\ref{tab:model_parameters}.

\begin{table*}
\centering
\caption{Best-fit leptonic and hadronic model parameters for the SED of PKS~0215+015 during the flaring epoch coincident with \textit{IceCube-220225A}, obtained within the one-zone framework of \citet{2013ApJ...768...54B}.}
\label{tab:model_parameters}
\begin{tabular}{lll}
\hline
Parameter & Symbol & Value \\
\hline
Bulk Lorentz factor & $\Gamma$ & 20 \\
Viewing angle (deg) & $\theta_{\rm obs}$ & 2.866 \\
Doppler factor & $\delta$ & 20 \\
Magnetic field strength (G) & $B$ & 1.0 \\
Emission region radius (cm) & $R$ & $5\times10^{15}$ \\
Minimum variability timescale (hr) & $\tau $ & 6.30 \\
Black hole mass ($M_{\odot}$) & $M_{\rm BH}$ & $5\times10^{8}$ \\
Accretion-disc luminosity (erg~s$^{-1}$) & $L_{\rm disc}$ & $1\times10^{46}$ \\
External radiation field temperature (K) & $T_{\rm ext}$ & $1.0\times10^{4}$ \\
External radiation field energy density (erg~cm$^{-3}$) & $u_{\rm ext}$ & $8.0\times10^{-2}$ \\
\hline
\multicolumn{3}{l}{\textit{Electron population}} \\
Minimum electron Lorentz factor & $\gamma_{e,\min}$ & $1.2\times10^{5}$ \\
Cutoff electron Lorentz factor & $\gamma_{e,c}$ & $1\times10^{6}$ \\
Electron injection spectral index & $q_{e}$ & 5.5 \\
Electron escape parameter & $\eta_{\rm esc}$ & 1.0 \\
Electron kinetic luminosity (erg~s$^{-1}$) & $L_{e}$ & $5.8\times10^{43}$ \\
\hline
\multicolumn{3}{l}{\textit{Proton population}} \\
Low-energy cutoff (GeV) & $E_{p,\min}$  & 1 \\
High-energy cutoff (GeV) & $E_{p,\max}$  & $1\times10^{10}$ \\
Proton spectral index & $\alpha_{p}$ & 2.0 \\
Proton kinetic luminosity (erg~s$^{-1}$) & $L_{p}$ & $4\times10^{45}$ \\
\hline
\multicolumn{3}{l}{\textit{Derived luminosity ratios}} \\
Magnetic-to-electron power ratio & $L_{B}/L_{e}$ & 0.647 \\
Magnetic-to-proton power ratio & $L_{B}/L_{p}$ & $9.38\times10^{-3}$ \\
Electron-to-proton power ratio & $L_{e}/L_{p}$ & $1.45\times10^{-2}$ \\
Proton-to-electron power ratio & $L_{p}/L_{e}$ & 69.0 \\
Proton-to-magnetic power ratio & $L_{p}/L_{B}$ & 107 \\
\hline
\end{tabular}
\end{table*}

\section{Discussion}
\label{sec:discussion}

The multi-wavelength behaviour of PKS~0215+015 around the epoch of \textit{IceCube-220225A} satisfies several of the criteria commonly invoked to assess a plausible blazar-neutrino association. The source is the only $\gamma$-ray emitter within the 90\% neutrino localisation region, lying $\sim$1.75~deg from the best-fit position, and was independently confirmed by \textit{Fermi}-LAT to be in a substantially elevated $\gamma$-ray state at the time of the alert, with a day-scale flux more than seven times, and a month-scale flux more than three times, its 4FGL-DR3 average \citep{2022ATel15243....1G}. This $\gamma$-ray enhancement was accompanied by an exceptional radio outburst, with flux densities exceeding 3~Jy, the highest recorded for this source in more than two decades \citep{2022ATel15247....1P}, and by VLBI evidence for a new superluminal jet component and associated polarization changes close in time to the neutrino arrival \citep{2026A&A...708A.326E}. The convergence of independent diagnostics, spanning radio, optical, $\gamma$-ray and parsec-scale jet kinematics, in the same $\sim$months-long window around MJD~59635 is difficult to attribute to chance, and constitutes the strongest qualitative evidence for a genuine physical connection between the flare and the neutrino detection.

Set against this favourable coincidence, however, our broadband SED modelling points to a more cautious, and physically more demanding, picture than the positional and temporal coincidence alone would suggest. The purely leptonic (synchrotron~+~SSC~+~EC) model reproduces the overall double-humped shape of the observed emission with a magnetic-to-electron luminosity ratio of $L_{B}/L_{e} \simeq 0.65$, close to equipartition. Achieving this fit, however, requires an electron energy distribution that is both narrow and injected at an unusually high minimum energy: the fitted population spans less than one decade in Lorentz factor, from $\gamma_{e,\min} = 1.2\times10^{5}$ to a cutoff at $\gamma_{e,c} = 1\times10^{6}$, rather than the broad, low-energy-anchored power law typically invoked in one-zone leptonic blazar models. Such a narrow, high-energy injection is not generic, and would require a correspondingly selective acceleration mechanism (e.g.\ magnetic reconnection or a similarly narrow-band process) to be physically justified rather than simply assumed. We further note that, while the leptonic model reproduces the broadband double-hump morphology, it does not fully reproduce the detailed shape of the \textit{Fermi}-LAT $\gamma$-ray spectrum.
We therefore regard the leptonic fit as a plausible working description of the flaring-epoch SED rather than as a fully satisfactory or unique one.

When a co-accelerated proton population is introduced to explore the maximum hadronic contribution allowed by the data, the resulting maximum allowed proton luminosity is roughly 69 times that of the electrons ($L_{p}/L_{e} \simeq 69$) and 107 times that of the magnetic field ($L_{p}/L_{B} \simeq 107$). This is a large deviation from equipartition, and reflects the baryon-loading problem commonly encountered when fitting neutrino blazars with lepto-hadronic models. Despite this large energy, the intrinsic inefficiency of proton synchrotron and photohadronic radiative channels keeps the hadronic contribution subdominant to the leptonic emission at every frequency.

To quantify the expected neutrino yield of this maximal hadronic scenario, we integrate the predicted neutrino flux against the IceCube effective area for the GFU-Bronze alert stream in the declination band containing PKS~0215+015, over three exposure windows. Over a single day, the maximal hadronic model predicts $N_{\nu} \simeq 1.3\times10^{-5}$; over the 10-day interval used to reconstruct the broadband SED (Section~\ref{sec:data}), it predicts $N_{\nu} \simeq 1.3\times10^{-4}$; and integrated over the full IceCube livetime to date ($\sim$15~yr), the same model predicts $N_{\nu} \simeq 7\times10^{-2}$. All three estimates are far below unity, consistent with the single, moderate-significance Bronze-tier alert actually observed rather than a statistically significant excess; a model with lower $L_{p}$, equally permitted by the SED, would predict a correspondingly lower rate in every case.

Taken together, the narrow electron distribution required by the leptonic fit and the extreme baryon loading permitted by the hadronic fit both point to the same underlying conclusion: a straightforward, generic one-zone model does not fully or comfortably explain this flaring episode, and we return to the implications of this tension for the proposed neutrino association in Section~\ref{sec:conclusion}.

An encouraging, model-independent cross-check comes from the agreement between two entirely independent estimates of the emission region size. The size derived from the observed short-term $\gamma$-ray variability, $\tau \simeq 6.61$~hr, together with the Doppler factor from the SED fit, gives $R \lesssim 5.17\times10^{15}$~cm (Eq.~\ref{eq:Rsize}), in close agreement with the value of $R = 5.0\times10^{15}$~cm required independently by the broadband SED fit itself. Because these two constraints are derived from entirely different data (a light-crossing argument applied to the $\gamma$-ray light curve versus a multi-band radiative-transfer fit), their consistency lends considerable confidence to the adopted jet geometry and Doppler factor, and by extension to the derived luminosities and baryon-loading ratios discussed above.

A further, more speculative but potentially important consideration concerns the long-standing ambiguity in the optical classification of PKS~0215+015. The source was originally identified as a BL~Lacertae object on the basis of its weak, featureless optical continuum, yet subsequent observations revealed prominent broad Ly$\alpha$ and C\,{\sc iv} emission lines \citep{1982ApJ...252..447G, 1982MNRAS.200.1091B, 1987AJ.....93..529F}, a transition characteristic of changing-look blazar behaviour \citep{2023NatAs...7.1282R}. Such changing-look episodes are frequently interpreted, at least in part, as manifestations of the masquerading BL~Lac scenario, in which an intrinsically FSRQ-type source with a genuine broad-line region is observed as a BL~Lac whenever its highly Doppler-boosted jet continuum outshines the comparatively faint disc/BLR emission, and reverts to an FSRQ-like appearance whenever the jet dims or the accretion power increases relative to it. If this interpretation applies to PKS~0215+015, then epochs of enhanced accretion activity, plausibly including the flaring state analysed here, could correspond to periods in which the BLR photon field is both intrinsically brighter and relatively more prominent against the jet continuum than assumed in our time-averaged, single-zone representation of $T_{\rm ext}$ and $u_{\rm ext}$. A denser external photon field of this kind would enhance the efficiency of photohadronic interactions without necessarily requiring the extreme proton kinetic luminosities implied by our present fit, offering a potential, physically motivated route to reconciling the source's favourable multi-wavelength and positional coincidence with \textit{IceCube-220225A} and the comparatively modest hadronic output found here. We stress that this connection is presented as a plausible implication of the source's classification history rather than as a result demonstrated by our modeling, and that testing it rigorously would require an epoch-resolved treatment of the BLR photon field rather than the constant external-radiation approximation adopted in this work.

More broadly, the pattern found here, a genuine and well-supported multi-wavelength coincidence accompanied by a leptonically-dominated SED and an extreme baryon loading requirement for any meaningful hadronic contribution, is not unique to PKS~0215+015 and recurs across a number of candidate neutrino-blazar associations. This recurrence highlights a general limitation of time-averaged, one-zone SED modeling for connecting individual neutrino alerts to specific blazar flares, and underscores the value of complementary approaches: epoch- and geometry-resolved external photon field modeling of the kind motivated above, multi-zone or stochastic emission models, and population-level statistical stacking analyses that do not depend on any single source providing an unambiguous hadronic signature.

\section{Conclusions}
\label{sec:conclusion}

We have presented a multi-wavelength study of the FSRQ PKS~0215+015, a candidate counterpart of the \textit{IceCube-220225A} neutrino event. Our main results are as follows.

\begin{enumerate}
\item PKS~0215+015 is the only $\gamma$-ray source within the 90\% localisation region of \textit{IceCube-220225A}, lying $\sim$1.75~deg from the best-fit neutrino position, and was independently confirmed to be in a substantially elevated $\gamma$-ray state at the neutrino epoch, together with a historic radio outburst and VLBI evidence for a new superluminal jet ejection and associated polarization changes over the same interval.

\item The broadband SED constructed for the flaring epoch is reproduced by a leptonic (synchrotron~+~SSC~+~EC) model with $L_{B}/L_{e} \simeq 0.65$, close to equipartition, though this requires an unusually narrow, high-energy electron distribution ($\gamma_{e,\min}=1.2\times10^{5}$, $q=5.5$) and does not fully capture the detailed \textit{Fermi}-LAT spectral shape. 
A hadronic contribution explored within the same framework corresponds to a maximum allowed proton kinetic luminosity roughly 69 times that of the electrons ($L_{p}/L_{e} \simeq 69$) and 107 times that of the magnetic field ($L_{p}/L_{B} \simeq 107$), yet remains radiatively subdominant to the leptonic emission at all frequencies, with a predicted neutrino event rate well below one detection over the lifetime of IceCube, consistent with the single, moderate-significance alert actually observed.

\item The emission region size derived independently from short-term $\gamma$-ray variability ($R \lesssim 5.17\times10^{15}$~cm) agrees closely with the value required by the SED fit ($R = 5.0\times10^{15}$~cm), providing a model-independent consistency check on the adopted jet geometry and Doppler factor.

\item The long-standing classification ambiguity of PKS~0215+015 between BL~Lac and FSRQ, and its possible interpretation as a masquerading BL~Lac / changing-look blazar, offers a plausible, though as yet untested, physical mechanism by which an epoch-dependent, enhanced BLR photon field could improve the prospects for photohadronic neutrino production beyond what is captured by our time-averaged external-radiation treatment.

\end{enumerate}

Further multi-wavelength and multimessenger monitoring of PKS~0215+015, combined with epoch-resolved modeling of its external radiation environment during periods of enhanced accretion activity, will help to clarify whether this source is a genuine, if energetically modest, neutrino emitter. More generally, continued systematic studies of candidate neutrino blazars such as PKS~0215+015 will inform the broader effort to identify the astrophysical sources of the diffuse high-energy neutrino flux, and to understand the physical conditions, whether compact emission regions, enhanced external photon fields, or extreme jet power, under which blazar jets become efficient hadronic accelerators.

\section*{Acknowledgments}
The authors acknowledge support by the South African Department of Science, Technology, and Innovation (DSTI) and the National Research Foundation (NRF) through funding for the South African Gamma-Ray Astronomy Programme (SA-GAMMA). This work makes use of archival $\gamma$-ray data from the Fermi Science Support Center (FSSC) and \textit{Swift}-XRT/UVOT data from the High Energy Astrophysics Science Archive Research Center (HEASARC).

\section*{Data Availability}

All the data used in this analysis are publicly available, and the results are incorporated in the paper.

\bibliographystyle{harv}
\bibliography{main}

\bio{}
\endbio

\bio{}
\endbio

\end{document}